\documentclass[final,5p,11pt,nopreprintline]{elsarticle}
\usepackage[T1]{fontenc}
\usepackage[utf8]{inputenc}
\usepackage{newtxtext,newtxmath}
\usepackage{microtype}
\usepackage{booktabs,array,tabularx}
\usepackage{amsmath}
\usepackage{xurl}
\usepackage[hidelinks]{hyperref}
\newcommand{\doi}[1]{doi:\href{https://doi.org/#1}{\nolinkurl{#1}}.}
\biboptions{numbers,sort&compress}
\makeatletter
\def\ps@pprintTitle{\let\@oddhead\@empty\let\@evenhead\@empty\def\@oddfoot{\hfil\thepage\hfil}\let\@evenfoot\@oddfoot}
\makeatother
\hypersetup{pdftitle={MedGeomBench-Core: Source-referenced regression testing of quantitative and spatial fidelity in medical image conversion},pdfauthor={Yuntao Wang},pdfsubject={Research preprint; revised 28 November 2026}}

\begin{document}
\begin{frontmatter}
\title{MedGeomBench-Core: Source-referenced regression testing of quantitative and spatial fidelity in medical image conversion}
\author{Yuntao Wang}
\begin{abstract}
\textbf{Background.} Medical image conversion can preserve numerical arrays while changing their assigned physical positions. Different checks address numerical equality, spatial representation, or reader compatibility, but these outcomes are often conflated.

\textbf{Purpose.} To develop and evaluate a source-referenced regression workflow with explicit criteria for quantitative and spatial preservation.

\textbf{Methods.} Eight retrospectively selected synthetic geometries, represented by 40 DICOM files, were reused in native-transport (72 attempts), array-interface (64 conversions), and reindexing (80 comparisons) panels. SimpleITK 2.5.8, VTK 9.6.2, and ImageIO 2.37.3 were evaluated using exact represented-value equality and a maximum coordinate-component difference of $10^{-6}$ mm. Separate scalar calculations, complete reruns, and installed-package execution were used for verification. A seven-threshold sensitivity analysis was performed post hoc.

\textbf{Results.} Native transport returned 52 quantity-exact arrays; 29 also met the spatial criterion, 23 did not, and 20 attempts were rejected. Array-interface conversion preserved all 64 arrays but only 22 spatial geometries. All 32 plain-array conversions changed geometry without conversion diagnostics. Forty reindexing comparisons preserved complete physical samples, including 32 legitimate reorders that did not preserve the original indexing. At 0.001 mm, diagnostic-silent discrepancies decreased from seven to zero for native transport and from 34 to 32 for array conversion. Installed-package reruns reproduced 660 arrays and 186 structured reports.

\textbf{Conclusions.} Numerical equality, diagnostic silence, and preservation of index geometry are not interchangeable. The workflow provides executable, bounded regression checks; existing methods matched the applicable reference decisions. Algorithmic superiority, clinical validity, and independent external reproducibility were not established.

\end{abstract}
\end{frontmatter}
\section{Introduction}
A medical image comprises both numerical samples and the spatial model assigning those samples to physical locations. DICOM specifies image-plane position, orientation, and pixel spacing, whereas NIfTI associates an array with spatial transforms \cite{dicom,nifti1}. Consequently, unchanged pixel values do not establish that a converted image occupies the same physical space. Conversely, a reader may reject an accurately represented file because the representation is unsupported. These are distinct outcomes, even when each occurs during an apparently simple conversion.

Conversion validation is well established. Yvernault et al. introduced an open multi-format resource containing DICOM images and validated conversions for testing transcoding tools \cite{yvernault2014}. Li et al. examined DICOM-to-NIfTI conversion challenges arising from manufacturers, acquisition types, and spatial interpretation \cite{li2016}. More recently, Rorden et al. provided 36 validation modules with corresponding NIfTI images and Brain Imaging Data Structure metadata \cite{rorden2025}.

Reproducible image-analysis software also has a substantial methodological foundation. SimpleITK exposes ITK functionality through a simplified programming interface \cite{lowekamp2013}; ITK's testing infrastructure and the SimpleITK notebook environment support reproducible development and executable image-analysis workflows \cite{mccormick2014,yaniv2018}. The present tool complements these resources through explicit comparison conditions, not a new conversion algorithm.

This study evaluated MedGeomBench-Core, a compact regression workflow that uses explicit source quantities and coordinates to distinguish serialization fidelity, reader compatibility, interface metadata propagation, and permissible index changes. The objective was not to introduce a new conversion algorithm or rank software generally. Three linked panels were curated from prior exploratory experiments, and a conservative file-reading interface was added after an audit identified a potential misinterpretation of object-level equality. The resulting scope is a reproducible software study using synthetic data, not a clinical validation study.

\section{Materials and methods}
\subsection{Study design and synthetic inputs}
Eight geometry configurations were retained from a parent experiment containing 24 series, with three mathematically equivalent value encodings per geometry. One encoding was selected for each geometry without modifying the DICOM bytes. Axial, sagittal, oblique, and deliberately large-origin configurations were each represented with orthogonal and sheared inter-slice progression. Each series contained five slices, giving 40 unique files. The large-origin cases were numerical stress conditions, not typical anatomical coordinates. All inputs were generated programmatically; no patient records, human participants, animals, or institutional clinical datasets were used.

Selection followed inspection of the parent results. Input hashes, code, and configuration were internally locked before each fresh execution, but the study was neither publicly preregistered nor evaluated on an independent holdout corpus. The same eight geometries were reused across panels. The 72, 64, and 80 evaluations therefore denote linked configuration tests, not independent samples. Counts are reported descriptively, without population prevalence estimates, confidence intervals, or significance testing.

\subsection{Software environment and source reference}
The tested environment was Linux x86\_64 with Python 3.13.5, NumPy 2.3.5, SimpleITK 2.5.8 (ITK 5.4), VTK 9.6.2, ImageIO 2.37.3, and pytest 9.0.2. DICOM discovery and sorting used the native SimpleITK interface, with float64 output and ForceOrthogonalDirectionOff to retain sheared geometry \cite{sitkseries}. Source quantities and voxel-center coordinates were checked against separately parsed DICOM declarations rather than assumed correct. Coordinate-system conversion between left-posterior-superior and right-anterior-superior conventions was fixed in advance; no registration, interpolation, or truth-informed permutation search was used.

\subsection{Native file-conversion panel}
Three writing configurations were evaluated: SimpleITK NIfTI-1, VTK NIfTI-1 with an explicitly supplied full sform, and VTK NIfTI-2 with the same sform policy. For VTK, the local image origin was zero, spacing was inherited from the source, and the transform accounted for the documented multiplication of sform columns by spacing \cite{vtk962}. The qform code was zero and the sform code was two. These were explicit interface choices, not a claim that all scanner-coordinate semantics or clinical metadata were retained.

Each of the 24 exported files was read using sform-aware VTK, strict SimpleITK, and SimpleITK with \nolinkurl{ITK_NIFTI_SFORM_PERMISSIVE=1} in a fresh child process, producing 72 attempts. VTK geometry was evaluated using the reader's exposed spatial transform and local spacing, not the image array's local coordinates alone \cite{vtk962}. VTK interface details were checked against documentation bundled with the installed 9.6.2 package, rather than a later nightly documentation page (Supplementary Methods S7). Arrays were compared in predetermined on-disk index order. All NIfTI fixtures in this panel had positive QFac; negative-qfac file interchange was not tested. The separate reindexing panel included reflections with negative-determinant direction matrices. A separate file-level check parsed the pixel payload and sform to distinguish changes already serialized in a file from changes introduced during reading.

\subsection{Array-interface panel and simple controls}
The installed ImageIO ITK adapter was used through its v2 and v3 application programming interfaces. Both APIs share the same legacy plugin and SimpleITK backend and were not treated as independent engines \cite{imageio}. For each retained source, four routes wrote both MetaImage (MHA) and NIfTI-1: direct SimpleITK object transport; ImageIO v2 volread/volwrite; ImageIO v3 imread/imwrite with is\_batch=False; and v3 array transport followed by CopyInformation from the source. The final route assumed unchanged arrays and indexing. CopyInformation transfers origin, spacing, and direction, not the complete metadata dictionary \cite{sitkimage}. The panel contained 64 conversions.

Two simple acceptance rules were assessed: a returned array with exact quantities, and the same requirement plus empty writer and reader standard-error streams. This diagnostic screen was not a classifier of warning meaning. Source-aware controls comprised SimpleITK IsSameImageGeometryAs with explicit 10\textsuperscript{-6} coordinate and direction tolerances, and an eight-corner coordinate test. These controls had access to the source image, whereas the diagnostic-only screen did not. Explicit metadata queries were recorded separately from conversion diagnostics. The adapter source was inspected before testing, making this a source-informed integration example rather than an unforeseen defect discovery.

\subsection{Preservation criteria and numerical comparison}
Quantitative preservation required exact equality of represented finite scalar values. For unchanged indexing, the spatial endpoint was the maximum absolute coordinate-component difference over all voxel centers. The threshold was 10\textsuperscript{-6} mm with zero relative tolerance. This deliberately strict numerical criterion was not a clinical tolerance. Nonfinite, shape-mismatched, or rejected outputs could not receive a fidelity pass; exceptions were retained without imputing arrays or geometry scores.

\begin{equation}
\begin{aligned}
E &= \max_{i,k}\left|x_{\mathrm{out},k}(i)-x_{\mathrm{ref},k}(i)\right|,\\
&\hspace{1.5em} E\leq 10^{-6}\ \mathrm{mm}.
\end{aligned}
\end{equation}
Here, i indexes voxel centers and k indexes the three physical-coordinate components. Same-index preservation required equal dimensions, represented values, and coordinates at each unchanged index. Complete physical-sample preservation permitted signed axis permutations, flips, and integer mappings but required all source samples, their positions, and their values to be retained. Crops and integer strides could match a source subset without satisfying complete equivalence. Common reference-frame identity and millimeter units were assumed, not inferred.

The reindexing panel applied ten fixed operations to each source, giving 80 comparisons. Native DICOMOrient canonicalization provided a simple source-aware control because it changes storage order while preserving sample locations \cite{sitkorient}. The corrected comparator avoided lossy integer-to-float equality tests and evaluated affine residuals using exact rational representations of the returned binary metadata \cite{numpy,python}. Earlier project defects and their impact are documented in Supplementary Methods S3. Supported inputs were finite, real, scalar, three-dimensional images with representations of at most 64 bits and at most 2,000,000 voxels. Arbitrary resampling and nonlinear mappings were outside scope.

\subsection{Conservative reading policy}
After an audit showed that permissive reading could make differently represented files appear equal as decoded objects, release 0.31.0rc1 added a conservative command-line boundary. Any inherited \nolinkurl{ITK_NIFTI_SFORM_PERMISSIVE} setting blocked automatic execution. Otherwise, a fresh child process set the variable to NO before importing the native library and enabled warning display. This string-based disabling behavior was checked experimentally and compared with the public tagged ITK source; the exact source commit of the installed binary was not independently reconstructed \cite{itk540}.

Any captured standard-output or standard-error bytes prevented automatic acceptance. Raw diagnostics, hashes, exceptions, timeouts, and process status were retained. Even a successful report explicitly stated that comparison concerned decoded image objects and did not certify on-disk fidelity. Entry files were hashed before and after reading, but sidecars and transient concurrent modifications were not authenticated. The boundary was neither a security sandbox nor a pre-decode memory limit. Its 23 command-line conditions and 35 additional tests were post-discovery regressions, not independent detection-performance estimates.

\subsection{Verification, installed execution, and sensitivity analysis}
Separate scalar implementations decoded relevant DICOM, NIfTI, and array bytes using only the Python standard library and calculated reference differences with exact rational arithmetic. They did not import the primary scorer, NumPy, VTK, or SimpleITK. Shared specifications and source declarations remained common dependencies. All checks were developed and executed within the same research process, so they were implementation cross-checks rather than independent-group validation.

Complete native runs were repeated in new directories. A wheel and source distribution were built, and installed console commands were executed outside the development directory in a separate virtual environment. Fixed host dependencies were shared; this did not constitute a different operating system or independently provisioned laboratory. Baseline execution success was kept separate from individual image fidelity. Original outputs, failed attempts, and amendment records were retained. Supplementary Methods S1-S4 describe the execution and correction history.

A post hoc sensitivity analysis re-evaluated existing native and array-interface outputs at 10\textsuperscript{-6}, 10\textsuperscript{-4}, 10\textsuperscript{-3}, 10\textsuperscript{-2}, 0.1, 1, and 10 mm. Exact quantitative equality remained mandatory, and rejected reads remained outside returned-output denominators. The panels were not pooled. Reindexing tolerances were not swept because changing lattice-matching tolerance changes the correspondence problem. The original primary criterion and classifications were retained.

\subsection{Artificial intelligence-assisted research}
ChatGPT (OpenAI) assisted study planning, literature review, protocol and code generation, debugging, execution orchestration, analysis, documentation, and manuscript preparation. Reported numbers were obtained from recorded local executions, not generated as unexecuted language-model results. Images were programmatically generated from recorded specifications, not by an image-generation model.

Complete historical model snapshot identifiers, generation settings, and exported prompt transcripts were unavailable. Reproduction therefore refers to the retained code and data, not regeneration of the original AI dialogue. Alternative implementations and automated checks arose from the same AI-assisted process and do not establish independent human review. Retained provenance and reporting limitations are detailed in Supplementary Methods S5. This research-method disclosure is distinct from assistance limited to manuscript wording.

\section{Results}
\subsection{Quantitative preservation and native reader compatibility}
All eight DICOM source reads matched the declared quantities and coordinates, and all 24 written payloads preserved their quantities. Of 72 downstream attempts, 52 returned quantity-exact arrays: 29 also met the spatial criterion and 23 did not. Twenty attempts raised caught native exceptions; none produced fabricated arrays or geometry scores. No native crashes or timeouts occurred in this panel (Table 1).

SimpleITK NIfTI-1 writing altered shear geometry in four configurations and emitted warnings. One additional orthogonal large-origin case exceeded the strict threshold because of limited precision. VTK NIfTI-1 retained shear but exceeded the threshold for two large-origin configurations; VTK NIfTI-2 retained the required precision in all eight files. VTK consistently interpreted the serialized sform in all 24 files. Strict SimpleITK rejected four sheared VTK NIfTI-1 files and all eight VTK NIfTI-2 files. Permissive reading returned the sheared NIfTI-1 files with warnings and altered coordinates; the largest source-relative component difference was 23.135216545 mm.

\begin{table*}[t]
\centering
\caption{Native NIfTI conversion outcomes for each writer-reader configuration.}
\label{tab:1}
\small
\renewcommand{\arraystretch}{1.14}
\setlength{\tabcolsep}{5pt}
\begin{tabularx}{\textwidth}{>{\raggedright\arraybackslash}p{0.22\textwidth}>{\raggedright\arraybackslash}Xccc}
\toprule
\textbf{Writer} & \textbf{Reader} & \textbf{Returned} & \textbf{Spatial pass} & \textbf{Exception} \\
\midrule
SimpleITK N1 & VTK & 8 & 3 & 0 \\
SimpleITK N1 & SimpleITK strict & 8 & 3 & 0 \\
SimpleITK N1 & SimpleITK permissive & 8 & 3 & 0 \\
VTK N1 / sform & VTK & 8 & 6 & 0 \\
VTK N1 / sform & SimpleITK strict & 4 & 3 & 4 \\
VTK N1 / sform & SimpleITK permissive & 8 & 3 & 0 \\
VTK N2 / sform & VTK & 8 & 8 & 0 \\
VTK N2 / sform & SimpleITK strict & 0 & 0 & 8 \\
VTK N2 / sform & SimpleITK permissive & 0 & 0 & 8 \\
\bottomrule
\end{tabularx}
\par\vspace{5pt}\noindent
\begin{minipage}{\textwidth}
\footnotesize Each row contains eight linked attempts using the same eight geometries. All returned arrays had exact quantities. Spatial pass requires a maximum coordinate-component difference of no more than $10^{-6}$ mm with zero relative tolerance. Exception denotes a caught native rejection or unsupported format, not an imputed output. N1, NIfTI-1; N2, NIfTI-2; sform, explicitly supplied full affine; strict, default SimpleITK reading; permissive, \nolinkurl{ITK_NIFTI_SFORM_PERMISSIVE=1} in a fresh process. These counts are not independent clinical samples.
\end{minipage}
\end{table*}
\subsection{Diagnostic screening and array-interface conversion}
Quantity equality alone accepted all 23 native spatial discrepancies. Requiring no writer or reader diagnostic output excluded all 16 returned shear-alteration records, leaving seven large-origin precision cases with discrepancies of approximately 0.000409330-0.000409578 mm (Table 2). These submicrometer differences exceeded the software threshold but do not demonstrate clinical harm.

\begin{table*}[t]
\centering
\caption{Simple acceptance policies applied to the native conversion panel.}
\label{tab:2}
\small
\renewcommand{\arraystretch}{1.14}
\setlength{\tabcolsep}{5pt}
\begin{tabularx}{\textwidth}{>{\raggedright\arraybackslash}X>{\centering\arraybackslash}p{0.10\textwidth}>{\centering\arraybackslash}p{0.22\textwidth}>{\centering\arraybackslash}p{0.20\textwidth}}
\toprule
\textbf{Acceptance rule} & \textbf{Accepted} & \textbf{Accepted spatial discrepancies} & \textbf{Largest discrepancy} \\
\midrule
Returned + exact quantities & 52 & 23 & 23.135217 mm \\
Above + no writer/reader stderr & 36 & 7 & 0.000409578 mm \\
\bottomrule
\end{tabularx}
\par\vspace{5pt}\noindent
\begin{minipage}{\textwidth}
\footnotesize Accepted spatial discrepancies are returned arrays that passed the stated policy but exceeded the primary $10^{-6}$ mm coordinate threshold. Largest discrepancy is the maximum absolute coordinate-component difference among outputs accepted by that policy. The absence of captured standard-error output is a diagnostic screen, not assurance that a file is correct. The seven discrepancies under the second rule were large-origin finite-precision conditions. No cutoff represents a clinical tolerance.
\end{minipage}
\end{table*}
All 64 array-interface conversions preserved quantities, but only 22 preserved source coordinates (Table 3). Each of the 32 plain-array conversions reset nondefault spatial information without conversion warnings or native standard-error output. Direct object transport and explicit geometry copying each preserved all eight MHA geometries and three of eight NIfTI-1 geometries. Their NIfTI discrepancies retained the known shear and large-origin effects; they were not additional independent discoveries.

The diagnostic-aware rule accepted 56 array conversions, including 34 spatial discrepancies. The existing SimpleITK geometry check accepted 22, all spatially concordant in this panel. Its decisions matched the corner-based and full-coordinate checks for all 64 conversions. No superiority of the custom check was demonstrated. For plain-array routes outside the large-origin family, maximum component differences ranged from 73.125 to 138.25 mm. These were constructed coordinate losses, not estimates of clinical error magnitude or frequency.

For every source, the v2 attached metadata dictionary was empty; each of the two explicit metadata-query APIs raised an exception, and v3 properties provided no spacing. These 16 unsupported-query outcomes were separate from conversion diagnostics because ordinary array transfer did not issue the queries. Respecting the explicit capability response would have prevented an unsupported metadata-preservation claim.

\begin{table}[t]
\centering
\caption{Spatially concordant outputs after image-object or array-interface conversion.}
\label{tab:3}
\small
\renewcommand{\arraystretch}{1.14}
\setlength{\tabcolsep}{5pt}
\begin{tabularx}{\columnwidth}{>{\raggedright\arraybackslash}Xcc}
\toprule
\textbf{Route} & \textbf{MHA} & \textbf{NIfTI-1} \\
\midrule
SimpleITK image & 8/8 & 3/8 \\
ImageIO v2 array & 0/8 & 0/8 \\
ImageIO v3 array & 0/8 & 0/8 \\
v3 array + CopyInformation & 8/8 & 3/8 \\
\bottomrule
\end{tabularx}
\par\vspace{5pt}\noindent
\begin{minipage}{\columnwidth}
\footnotesize Entries are spatially concordant outputs divided by eight linked source geometries. All 64 arrays preserved quantities exactly. MHA, MetaImage; NIfTI-1, Neuroimaging Informatics Technology Initiative version 1. The explicit CopyInformation control assumed unchanged indexing and copied origin, spacing, and direction rather than all clinical metadata. ImageIO v2 and v3 shared the same plugin and backend.
\end{minipage}
\end{table}
\subsection{Reindexing and partial coverage}
All 40 identity or legitimate-reindexing comparisons preserved complete physical samples. Only the eight identities also preserved unchanged indexing; the other 32 correctly altered storage correspondence. Sixteen crops or strides preserved a subset, 16 comparisons had changed values, and eight shifted-origin cases were off the source lattice. Native canonicalization and the physical-sample check agreed on all 80 complete-equivalence decisions. These outcomes characterize preservation, not the correctness of an intentionally different processing task.

\subsection{Threshold sensitivity}
The two retained coordinate references agreed on all 952 configuration-threshold decisions. At 0.001 mm, all seven diagnostic-silent native discrepancies were within threshold, whereas the 16 larger native discrepancies remained. In the array-interface panel, 32 diagnostic-silent outputs remained discrepant even at 10 mm (Table 4). Rejected reads remained excluded from returned-output denominators. None of these descriptive cutoffs was selected or validated as a clinical tolerance.

\begin{table*}[t]
\centering
\caption{Post hoc sensitivity of spatial-discrepancy counts to the coordinate threshold.}
\label{tab:4}
\small
\renewcommand{\arraystretch}{1.14}
\setlength{\tabcolsep}{5pt}
\begin{tabularx}{\textwidth}{>{\raggedright\arraybackslash}p{0.15\textwidth}*{4}{>{\centering\arraybackslash}X}}
\toprule
\textbf{Threshold (mm)} & \textbf{Native mismatch /52} & \textbf{Silent native mismatch /36} & \textbf{Array mismatch /64} & \textbf{Silent array mismatch /56} \\
\midrule
1e-06 & 23 & 7 & 42 & 34 \\
0.0001 & 23 & 7 & 42 & 34 \\
0.001 & 16 & 0 & 40 & 32 \\
0.01 & 16 & 0 & 40 & 32 \\
0.1 & 16 & 0 & 40 & 32 \\
1 & 7 & 0 & 34 & 32 \\
10 & 2 & 0 & 32 & 32 \\
\bottomrule
\end{tabularx}
\par\vspace{5pt}\noindent
\begin{minipage}{\textwidth}
\footnotesize Native counts use the 52 returned outputs; 20 rejected or unsupported reads are excluded at every threshold. Silent native counts use the 36 returned outputs with no captured writer or reader diagnostics. Array counts use 64 outputs, of which 56 had no captured diagnostics. Exact quantitative equality remains mandatory. The original endpoint was not changed by this post hoc analysis, and none of the listed thresholds was clinically validated.
\end{minipage}
\end{table*}
\subsection{Reproducibility and corrective-release behavior}
The corrected post-audit retained all 72 transport, 64 array-interface, and 80 reindexing classifications. Seventy-nine diagnostic values differed from earlier calculations, with a maximum absolute change of $1.8189894035458565\times10^{-12}$ mm. These changes reflected exact arithmetic and the explicitly different source-coordinate representations, not altered primary endpoints. Both diagnostic definitions and values were retained.

The final installed package passed 87 tests: 52 inherited preflight tests and 35 reading-boundary tests. Two complete runs reproduced all 660 arrays and 186 structured or scientific JSON reports. Of 974 compared files, 958 were byte-identical; the remaining 16 raw logs differed in run paths or native object addresses. Both raw logs and separately labeled normalized comparisons were retained. A standard-library recheck supported 47,850 quantity comparisons and 143,550 coordinate-component comparisons against source DICOM declarations.

On the retained challenge, the earlier console returned success under both object-level contracts after permissive reading, despite a 0.5 mm difference between the files' serialized corner coordinates and the presence of native warnings. The revised command blocked the inherited setting before decoding. With no inherited setting, strict native reading rejected the nonorthogonal challenge. All 23 targeted command-line checks matched their stated expectations. These included conservative refusals and cannot be interpreted as sensitivity or specificity. Preliminary implementation and packaging failures, their amendments, and final reruns are reported in Supplementary Methods S3-S4.

\section{Discussion}
This study demonstrates why quantitative preservation, file representation, reader compatibility, and index correspondence should be assessed separately. Every successfully returned array in the native panel was quantity-exact, yet some assigned different positions to those quantities. Conversely, some accurately represented outputs were rejected. Array-interface conversion exposed a further failure mode: spatial information could be discarded before the writer was called, leaving a valid default-grid image and no conversion warning.

The format and the implementation must also be distinguished. NIfTI-1 sform permits general affine geometry; finite field precision and a particular writer are separate limitations \cite{nifti1}. NIfTI-2 increases the precision of several spatial fields, but does not force downstream readers to support the format \cite{nifti2}. The tested strict and permissive behaviors have existing upstream context \cite{itkissue}. They are included as reproducible regression cases rather than novel defects, and neither a higher-precision export nor a warning-bearing successful read is a universally suitable conversion policy.

The simple controls limit the strength of any algorithmic claim. Diagnostic screening excluded all larger returned geometric changes in the native panel. At 0.001 mm, its seven residual precision discrepancies disappeared. In the array panel, an existing source-aware geometry interface matched the full reference decisions; in the reindexing panel, native canonicalization matched complete physical equivalence. The useful contribution is therefore a documented set of comparison requirements, provenance-preserving execution, and interpretable outcomes. It is not evidence that a novel detector outperforms established methods. Source-aware methods also receive information unavailable to warning-only rules.

Legitimate processing should not be mislabeled as failure merely because it changes storage or coverage. A correct axis permutation need not preserve the original index grid, and a correct crop intentionally removes samples. Conversely, copying a source geometry onto a rearranged array may produce a plausible header attached to incorrect values. Explicitly specifying unchanged indexing or complete physical-sample equivalence avoids a generic pass/fail label with an unclear meaning. Neither condition establishes patient identity, a shared reference-frame identity, or correct absolute placement.

The audit of the checker further showed that object-level equivalence does not certify file-level fidelity. A reader can alter metadata before comparison, and warnings may be lost by applications that inspect only an exit code. The conservative boundary makes this limitation machine-readable and prevents automatic acceptance after captured diagnostics or inherited permissive settings. It does not repair the reader, exclude unreported alterations, or provide a security boundary. Refusals of potentially acceptable files are a coverage cost, not successfully detected defects.

Several limitations restrict generalization. The eight geometries were chosen after earlier outcomes were known, and all panels reuse the same 40 files. Equivalent encodings, repeated configurations, and reruns are not independent observations. No clinical error frequency, patient outcome, population performance, or software-wide ranking can be inferred. The large-origin tests were deliberately nonclinical numerical stresses, and the 10\textsuperscript{-6} mm threshold must not be presented as clinical acceptability. Compressed or multiframe DICOM and negative-qfac NIfTI reader-writer interchange were not evaluated. The reindexing panel did include reflections with negative-determinant direction matrices, which do not establish support for negative-qfac file interchange. Arbitrary qform/sform disagreement, registration, general interpolation, missing-source completeness, and preservation of all clinical metadata or unit labels remained outside this core evaluation.

Only one host environment and fixed package versions were tested; the exact ITK build-source commit was not independently reconstructed. Separate algorithms developed within one AI-assisted project and a new virtual environment on the same host are not independent scientific replication. Existing conversion resources provide broader manufacturer and acquisition coverage \cite{yvernault2014,li2016,rorden2025}. MedGeomBench-Core separates quantity preservation, serialized geometry, reader acceptance, metadata transfer, and reindexing conditions; it neither replaces their coverage nor establishes a superior detector. Its incremental benefit in a maintained external application, independent-group reproducibility, and production readiness remain untested.

\section{Conclusions}
MedGeomBench-Core provides a bounded, executable framework for source-referenced regression testing of medical image conversion. It distinguishes unchanged quantities from preserved spatial coordinates, unsupported representations, and legitimate changes in indexing or coverage. Repeated native execution and alternative scalar calculations support the reported results, but do not establish algorithmic superiority or clinical validity. The appropriate use is reviewable software regression testing under explicitly stated assumptions, not unconditional certification of images or files.

\section*{Funding}
This research received no funding.
\section*{Data and code availability}
The author retains release 0.31.0rc1, its code, 40 synthetic DICOM files, hashes, raw outputs, diagnostics, audits, amendments, and reproduction instructions. No public repository, archival DOI, or project license had been established at the date of this preprint. The compact software distribution excludes native dependencies; a separate private offline backup includes third-party installers and an environment snapshot and is not the proposed public release. Redistribution requires separate review (Supplementary Methods S6). The unexecuted NiBabel and Ni-Dataset comparison is outside the reported scope.

\section*{Ethics statement}
All study inputs were synthetic. No human participants, animal experiments, patient records, or identifiable clinical data were involved.

\ifdefined\ArticleOnly
\else
\clearpage
\section*{Supplementary methods}
\section*{S1. Experimental units, configurations, and reference definitions}
The final corpus contains eight geometry configurations and 40 unique DICOM files. It was obtained by retaining one encoding, enc0, for each of eight geometries from a parent panel of 24 series. The selected bytes were not modified. The geometry families are axial, sagittal, oblique, and numerical large-origin stress, each with orthogonal and sheared slice progression. Each series contains five slices. The 72 native-conversion attempts, 64 array-interface conversions, and 80 reindexing comparisons reuse these inputs and must not be summed as independent subjects, acquisition series, or clinical cases.

The native-conversion panel combines three writers and three reader policies for each geometry. The array panel combines four routes and two output formats. The reindexing panel includes identity, legitimate flips and permutations, native orientation canonicalization, integer cropping and striding, deliberate quantity changes, incorrect geometry copying after rearrangement, and shifted-origin conditions. Full operation definitions, source hashes, and raw outputs are retained in the research archive.

The primary coordinate reference is the geometry declared by DICOM and the independently parsed serialized affine. A later mandatory post-audit additionally uses the represented native source image or saved source affine. Those coefficients can differ slightly from the decimal DICOM declarations. The two references are explicitly distinguished; diagnostic values are not silently substituted. Their seven-threshold classifications agreed in all 952 configuration-threshold evaluations.

\section*{S2. Alternative implementations and replay accounting}
The native-transport scalar audit checked 224,400 quantities or coordinate components and 1,432 score or header fields per run. The array-interface audit checked 224,400 quantities or coordinate components, 384 score fields, and eight source orderings. The reindexing audit checked 127,440 numerical or coordinate components and 536 fields. These numbers describe related implementation checks, not independent samples. The alternatives did not import the primary scorer or native image libraries, but shared source declarations and specifications, and were developed within the same research process.

Individual panels initially passed 32, 15, and 21 pre-execution tests, respectively. Their final stage counts were 43, 32, and 37 after adding post-result regression and evidence checks. These counts overlap with later bundled tests and must not be added to obtain an independent test total. In the integrated run, 168 code, configuration, and input files were locked before execution. The corrective release retained 173 embedded historical study files unchanged.

Across final complete runs, all 220 NPZ archives containing 660 arrays and all 186 structured or scientific JSON files agreed. Of 974 compared files, 958 were byte-identical. Sixteen raw diagnostic logs differed in native object addresses or output-directory spelling. Original streams were kept alongside a separately labeled normalization comparison. No exception was deleted or replaced by a fabricated output. A final standard-library recheck supported 47,850 quantities and 143,550 coordinate components against DICOM declarations.

\section*{S3. Corrections to the project comparator}
Two defects were reproduced in an earlier comparator. First, promotion during mixed-type array equality could treat integer 9,007,199,254,740,993 as equal to float64 9,007,199,254,740,992. Second, direct subtraction of coordinates near a common 10\textsuperscript{16} mm origin could hide a 0.3 mm residual caused by an inter-sample spacing change. The same-index check already rejected the latter case; the physical-sample path required correction. These are project implementation errors and nonclinical numerical stresses, not newly discovered upstream defects.

The corrected comparator uses exact represented-value semantics without a lossy integer-to-float promotion, and exact rational arithmetic for residuals of the binary metadata returned by the image API. These calculations cannot restore precision already lost during serialization or reading. The audit is limited to finite real scalar three-dimensional images with at most 64-bit representations and 2,000,000 voxels per input. Common physical units and reference-frame identity are assumptions.

The integrated correction changed no primary classification in the three panels. Seventy-nine diagnostic fields changed: 35 native-transport, 20 array-interface, and 24 reindexing fields. The largest old-to-new absolute difference was $1.8189894035458565\times10^{-12}$ mm. Both values and their reference definitions were retained. Passing a finite test suite is not a proof for arbitrary types, dimensions, mappings, or files.

\section*{S4. Reading-interface challenge and corrective release}
In the retained challenge, VTK wrote two quantity-identical NIfTI files whose sforms differed by a maximum corner-coordinate component of 0.5 mm. Under permissive native reading, an earlier installed command reported both decoded-object contracts as satisfied while the reader emitted warnings. This exposed ambiguity in interpreting a successful object comparison as file preservation. The original documentation had limited the comparator to decoded objects; the revision made that limitation explicit in the machine-readable result and added conservative reading policy.

The corrected command refuses any inherited \nolinkurl{ITK_NIFTI_SFORM_PERMISSIVE} variable. With no inherited setting, a fresh child process sets the explicit disabling token NO before importing SimpleITK and captures diagnostic streams. Native probes found that absent, NO, OFF, and FALSE settings rejected the nonorthogonal challenge, whereas 0, 1, and an empty string enabled permissive handling. An initial local attempt had incorrectly used 0. Its source, output, and amendment were retained, and the final frozen runs used NO.

Automatic acceptance is withheld whenever captured standard output or standard error is nonempty. Reports retain bytes, decoded diagnostic text, hashes, process status, exceptions, and timeouts. A successful report states \nolinkurl{comparison_basis=decoded_image_objects_only} and \nolinkurl{on_disk_fidelity_certified=false}. Conservatively blocking a configuration or diagnostic may reject an otherwise acceptable object; such outcomes are a coverage cost and are not counted as detected defects. Entry-file hashes do not authenticate sidecars or transient concurrent modifications. The process boundary is not a security sandbox or a pre-decode memory guarantee.

The final installed package passed 87 tests: 52 inherited study preflight tests and 35 additional boundary tests. Twenty-three actual command-line checks matched their declared expectations, covering normal images, legitimate flips under different contracts, crops, changed quantities, numerical edges, unsupported inputs, diagnostic-bearing returns, inherited settings, timeout, missing files, and protection of existing reports. They were designed after discovery and are not a blinded performance test.

A separate packaging assertion found four generated bytecode files in an intermediate wheel. A grouped launch command initially continued despite that failure, so an early run was retained as provisional rather than relabeled as the final run. The package exclusion rules and fail-fast launch were corrected, then the wheel was rebuilt, installed, locked, and tested again. The frozen numerical code and fixtures were not altered to obtain agreement. A virtual environment had also initially lacked pytest; explicit host-dependency discovery was corrected before the corresponding scientific execution.

\section*{S5. AI-assisted workflow and provenance}
ChatGPT assisted planning, literature inspection, protocol and code generation, debugging, execution orchestration, analysis, documentation, and manuscript writing. All experimental arrays were created programmatically. Numerical tables arose from retained executions and were checked against structured outputs. Code locks, source and input hashes, raw outputs, unsuccessful attempts, amendments, and final evidence records remain in the local research archives.

The retained records identify November 2026 as the development period and November 28, 2026 as the corrective integration date. Complete historical model snapshot identifiers, generation parameters, and exported prompt transcripts are unavailable. Re-execution of the fixed code does not require regeneration of the AI dialogue; reconstruction of that dialogue is not claimed. Tests and alternative implementations from this process are not independent human review.

\section*{S6. Distribution and interpretation of command outcomes}
The research archive includes a locally built and tested wheel and source distribution for release 0.31.0rc1. Console commands include \nolinkurl{medgeombench-run}, \nolinkurl{medgeombench-audit}, \nolinkurl{medgeombench-selftest}, and \nolinkurl{medgeombench-examples}. Installed execution was performed outside the development directory. The same Linux host and pinned native dependencies were shared, so this tests packaging and path independence, not cross-platform or independent-laboratory replication.

A zero exit from \nolinkurl{medgeombench-run} means the fixed study completed and matched its recorded regression baseline, including expected rejections and discrepancies. A zero exit from \nolinkurl{medgeombench-audit} means only that the chosen decoded-object comparison was satisfied under the guarded reader policy without captured diagnostics. Neither outcome is a universal file certificate. Full clinical metadata, semantic units, source completeness, patient identity, and absolute placement are not established. A public repository, archival DOI, and finalized project license had not been created at the date of this preprint. The compact software distribution excludes native dependencies. A separate private preservation backup includes third-party installation archives and a same-platform environment snapshot; it is not the proposed public software release. Public redistribution of those materials requires a separate license review.

\section*{S7. Technical source and version traceability}
The references distinguish a format specification, package documentation, and contextual upstream discussion from the exact binary tested. The DICOM Image Plane Module page identified PS3.3 edition 2026e when checked on 28 November 2026. The NIfTI references use the official field definitions and the approved 2011 NIfTI-2 description rather than an unversioned development header. These sources support format properties; they do not establish any reader's compliance.

VTK-specific sform behavior was checked against Python method documentation distributed with the installed version 9.6.2 package. The checked methods were vtkNIFTIImageWriter.SetSFormMatrix, SetQFormMatrix, and SetNIFTIVersion, and vtkNIFTIImageReader.GetSFormMatrix, GetQFormMatrix, and GetQFac. Version identity was obtained from both the package metadata and vtkVersion.GetVTKVersion(). In particular, the writer documentation specifies multiplication of the first three sform columns by voxel spacing; the reader documentation specifies factoring local spacing out of its returned spatial matrix. Dated documentation from a later nightly build was not used as exact-version evidence.

SimpleITK ImageSeriesReader and Image interface references are fixed at tag v2.5.8. The DICOMOrient reference is explicitly identified as version-2.5 documentation rather than falsely labeled as an exact patch-specific page. The installed SimpleITK package was 2.5.8. Public ITK v5.4.0 source was used as behavior context only: the exact source commit used to build the installed ITK binary was not independently reconstructed. NumPy and Python documentation identify the 2.3 and 3.13 series, respectively, while the actual tested patch versions are reported in the main Methods.

The bibliography separates six research articles from the necessary format and implementation references. General packaging instructions and publisher policies are not used as scientific evidence of novelty. All DOI links, citation keys, table entries, and version statements were checked during the editorial revision. No new native experiments or numerical endpoints were introduced by that revision.

\fi

\begin{thebibliography}{99}
\small
\interlinepenalty=5000
\setlength{\itemsep}{3pt}
\bibitem{dicom}
DICOM Standards Committee. \href{https://dicom.nema.org/medical/dicom/current/output/chtml/part03/sect_C.7.6.2.html}{Digital Imaging and Communications in Medicine (DICOM), PS3.3: Information Object Definitions}. 2026e, Section C.7.6.2, Image Plane Module. Accessed 2026-11-28.

\bibitem{nifti1}
NIfTI Data Format Working Group. \href{https://nifti.nimh.nih.gov/nifti-1/documentation/nifti1fields.html}{NIfTI-1 header fields and spatial transforms}. NIfTI-1 specification. Accessed 2026-11-28.

\bibitem{yvernault2014}
Yvernault BC, Theobald CD Jr, Smith JC, Villalta V, Zald DH, Landman BA. Validating DICOM transcoding with an open multi-format resource. \textit{Neuroinformatics}. 2014;12(4):615--617. \doi{10.1007/s12021-014-9230-9}

\bibitem{li2016}
Li X, Morgan PS, Ashburner J, Smith J, Rorden C. The first step for neuroimaging data analysis: DICOM to NIfTI conversion. \textit{J Neurosci Methods}. 2016;264:47--56. \doi{10.1016/j.jneumeth.2016.03.001}

\bibitem{rorden2025}
Rorden C, B\'eranger B, Cheng H, Clemence M, Debacker C, Fernandez B, et al. DICOM datasets for reproducible neuroimaging research across manufacturers and software versions. \textit{Sci Data}. 2025;12:1168. \doi{10.1038/s41597-025-05503-w}

\bibitem{lowekamp2013}
Lowekamp BC, Chen DT, Ib\'a\~nez L, Blezek D. The design of SimpleITK. \textit{Front Neuroinform}. 2013;7:45. \doi{10.3389/fninf.2013.00045}

\bibitem{mccormick2014}
McCormick M, Liu X, Jomier J, Marion C, Ibanez L. ITK: enabling reproducible research and open science. \textit{Front Neuroinform}. 2014;8:13. \doi{10.3389/fninf.2014.00013}

\bibitem{yaniv2018}
Yaniv Z, Lowekamp BC, Johnson HJ, Beare R. SimpleITK image-analysis notebooks: a collaborative environment for education and reproducible research. \textit{J Digit Imaging}. 2018;31(3):290--303. \doi{10.1007/s10278-017-0037-8}

\bibitem{sitkseries}
SimpleITK Developers. \href{https://raw.githubusercontent.com/SimpleITK/SimpleITK/v2.5.8/Code/IO/include/sitkImageSeriesReader.h}{ImageSeriesReader interface}. SimpleITK 2.5.8 [source code]. Accessed 2026-11-28.

\bibitem{vtk962}
VTK Developers. \href{https://pypi.org/project/vtk/9.6.2/}{Visualization Toolkit: vtkNIFTIImageWriter and vtkNIFTIImageReader interfaces}. Version 9.6.2; bundled Python API documentation. Accessed 2026-11-28.

\bibitem{imageio}
ImageIO Developers. \href{https://imageio.readthedocs.io/en/v2.37.3/reference/core_v3.html}{Core API v3 and metadata interfaces}. ImageIO 2.37.3. Accessed 2026-11-28.

\bibitem{sitkimage}
SimpleITK Developers. \href{https://raw.githubusercontent.com/SimpleITK/SimpleITK/v2.5.8/Code/Common/include/sitkImage.h}{Image interface: CopyInformation and IsSameImageGeometryAs}. SimpleITK 2.5.8 [source code]. Accessed 2026-11-28.

\bibitem{sitkorient}
SimpleITK Developers. \href{https://simpleitk.org/doxygen/v2_5/html/classitk_1_1simple_1_1DICOMOrientImageFilter.html}{DICOMOrientImageFilter class reference}. Version 2.5 documentation; tested package 2.5.8. Accessed 2026-11-28.

\bibitem{numpy}
NumPy Developers. \href{https://numpy.org/doc/2.3/reference/arrays.promotion.html}{Data type promotion in NumPy}. Version 2.3 manual. Accessed 2026-11-28.

\bibitem{python}
Python Software Foundation. \href{https://docs.python.org/3.13/library/stdtypes.html#numeric-types-int-float-complex}{Built-in types: numeric types and comparisons}. Python 3.13 documentation. Accessed 2026-11-28.

\bibitem{itk540}
Insight Software Consortium. \href{https://raw.githubusercontent.com/InsightSoftwareConsortium/ITK/v5.4.0/Modules/IO/NIFTI/src/itkNiftiImageIO.cxx}{NiftiImageIO implementation: itkNiftiImageIO.cxx}. ITK 5.4.0 [source code]. Accessed 2026-11-28.

\bibitem{nifti2}
NIfTI Data Format Working Group. \href{https://www.nitrc.org/forum/message.php?msg_id=3738}{NIfTI-2: a 64-bit update of NIfTI-1 (approved)}. 2011; NITRC message 3738. Accessed 2026-11-28.

\bibitem{itkissue}
Insight Software Consortium. \href{https://github.com/InsightSoftwareConsortium/ITK/issues/3994}{NIfTI files with non-orthonormal directional cosines should be read and the error changed to a warning}. ITK issue 3994; opened 4 April 2023. Accessed 2026-11-28.

\end{thebibliography}
\end{document}